\documentclass[showpacs,amsmath,amssymb,pre]{revtex4}
\usepackage{bm}
\usepackage[dvips]{graphicx}
\newcommand{\D}{\text{d}}
\newcommand{\lrho}{\tilde{\rho}}

\graphicspath{{../fig/}}

\begin{document}

\title{Structure of a Liquid Crystalline Fluid around a Macroparticle: Density
  Functional Theory Study}

\author{David L. Cheung}
\email{david.cheung@warwick.ac.uk}
\author{Michael P. Allen}
\email{m.p.allen@warwick.ac.uk}
\affiliation{Department of Physics and Centre for Scientific Computing,
  University of Warwick, Coventry, CV4 7AL, UK}

\begin{abstract}
  The structure of a molecular liquid, in both the nematic liquid crystalline
  and isotropic phases, around a cylindrical macroparticle, is studied using
  density functional theory. In the nematic phase the structure of the fluid
  is highly anisotropic with respect to the director, in agreement with
  results from simulation and phenomenological theories.  On going into the
  isotropic phase the structure becomes rotationally invariant around the
  macroparticle with an oriented layer at the surface.
\end{abstract}
\maketitle

\section{Introduction}

Solid particles, both spherical and non-spherical, dispersed in a liquid crystal
(LC) host comprise an interesting class of novel
materials \cite{poulin.p:1999.a,stark.h:2001.a}. In orientationally ordered phases of the
host, the introduction of solid particles deforms the director field, leading to
long-range interactions between the particles and effects such as chaining or
the formation of soft solids.  Colloidal dispersions in liquid crystals have a
wide range of applications \cite{russel.wb:1989.a} and have recently
attracted a great deal of interest \cite{jpcm_lccolloid_specialissue}.

Experimental techniques such as atomic force microscopy or confocal microscopy
may be used to study LC-colloid dispersions; simulations and theory have also
been applied. Simulations have been used to study the ordering of LC molecules
around one \cite{billeter.jl:2000.a,andrienko.d:2001.a} or two macroparticles
\cite{albarwani.ms:2004.a}, and LC dispersions in confined geometries
\cite{kim.eb:2002.a}. Phenomenological theories such as Landau-de Gennes
\cite{borstnik.a:1999.a,patricio.p:2002.a,tasinkevych.m:2002.a,andrienko.d:2004.a}
or Frank elastic \cite{stark.h:1999.a,yamamoto.r:2004.a} theory have also been
used. These approaches however have their limitations: simulation is
computationally expensive, while the aforementioned phenomenological theories
require, often poorly known, parameters, and are incapable of accounting for
spatial variation in the density (and, in the case of elastic theory,
variation in the order parameter).

One popular theoretical method that may be applied to this type of problem is
density functional theory (DFT) \cite{harnau.l:to_appear}. Unlike Landau-de
Gennes or elastic theory, this is capable of accounting for spatial variation
in the density and order parameter and, in the form used here, requires
knowledge only of the interaction potential between the molecules in the
fluid.  In this case DFT at the level of the Onsager second virial
approximation \cite{onsager.l:1949.a} is applied to the case of a single,
infinitely long, cylindrical macroparticle in a LC host. As the system is
homogeneous along the length of the cylinder, two dimensions are sufficient to
represent the spatial dependence of the density.

\section{Theory}

For a system of uniaxial molecules the grand potential can be written
as \cite{hansen.jp:1986.a}
\begin{equation}
  \beta\Omega\left[\rho(\bm{r},\bm{u})\right] =
  \beta F_\text{id}\left[\rho(\bm{r},\bm{u})\right]+
  \beta F_\text{ex}\left[\rho(\bm{r},\bm{u})\right]
  + \beta\int\;\D\bm{r} \D\bm{u}\; \bigl( V_\text{ext}(\bm{r},\bm{u})-\mu \bigr) \rho(\bm{r},\bm{u}) \;,
\end{equation}
where $\rho(\bm{r},\bm{u})$ is the position- and orientation-dependent single particle
density, $V_\text{ext}(\bm{r},\bm{u})$ is the external potential, $\mu$ is the
chemical potential, and $\beta=1/k_BT$. $F_\text{id}\left[\rho(\bm{r},\bm{u})\right]$
and $F_\text{ex}\left[\rho(\bm{r},\bm{u})\right]$ are the ideal and excess free
energies, respectively. $\bm{r}$ is the position vector and $\bm{u}$ is the
orientation vector. The ideal free energy is given by
\begin{equation}
  \beta F_\text{id}\left[\rho(\bm{r},\bm{u})\right]=
  \int\;\D\bm{r} \D\bm{u}\; \rho(\bm{r},\bm{u})\bigl\{
    \log\rho(\bm{r},\bm{u})-1\bigr\} \;.
\end{equation}
The exact form of the excess free energy is generally unknown. Here we
employ the Onsager approximation \cite{onsager.l:1949.a} 
\begin{equation}
\label{eqn:onsager}
  \beta F_\text{ex}\left[\rho(\bm{r},\bm{u})\right] =
  -\frac{1}{2}\int\;\D\bm{r}_1\D\bm{r}_2\D\bm{u}_1\D\bm{u}_2\; 
  f(\bm{r}_{12},\bm{u}_1,\bm{u}_2)\rho(\bm{r}_1,\bm{u}_1)\rho(\bm{r}_2,\bm{u}_2) \;,
\end{equation}
where $f(\bm{r}_{12},\bm{u}_1,\bm{u}_2)=\exp\bigl\{-\beta
V(\bm{r}_{12},\bm{u}_1,\bm{u}_2)\bigr\}-1$, the Mayer function,
$\bm{r}_{12}=\bm{r}_1-\bm{r}_2$ and $V(\bm{r}_{12},\bm{u}_1,\bm{u}_2)$ is the
molecular pair interaction potential.  The liquid crystal here is modelled as a
fluid of prolate hard ellipsoids of elongation $e=a/b=5$; $a$ is the length of
the symmetry, or major, axis and we will use the minor axis $b=1$ as a unit of
length.  When two molecules overlap, $V=\infty$ and $f=-1$; for a
non-overlapping pair, $V=0$ and $f=0$.  The approximation of
eqn~\eqref{eqn:onsager} corresponds to truncating the virial expansion after the
pair term.  While Onsager theory is exact only in the limit of infinite
elongation, it has been used to study the anchoring of ellipsoids of this
elongation near solid substrates \cite{chrzanowska.a:2001.a,teixeira.pic:2004.a}
and has been found to give results in qualitative agreement with simulation.

The external potential, representing a single cylindrical macroparticle of radius
$R$ oriented along the $y$ axis, is given by
\begin{equation}
  V_\text{ext}(\bm{r},\bm{u})=
  V_\text{ext}(\bm{s},\bm{u})=\begin{cases}
  \frac{1}{2}V_0\bigl[\textrm{tanh}(b/w)-\textrm{tanh}(-b/w)\bigr] & s-R<-b \\
  \frac{1}{2}V_0\left[\textrm{tanh}\Bigl(\dfrac{R-s}{w}\Bigr)-\textrm{tanh}(-b/w)
    \right]
  & |s-R| < b \\
      0 & s-R>b 
      \end{cases}
\end{equation}
where $\bm{s}=(x,z)$, $s$ = $|\bm{s}|$,
$V_0=50k_BT$ and $w=b/5$. This represents a sharply varying repulsive
potential acting on the ellipsoid centres of mass. It excludes the molecules
from the cylinder and gives rise to homeotropic (normal) anchoring at the surface.

As before
\cite{allen.mp:1999.a,allen.mp:2000.c,andrienko.d:2002.a,cheung.dl:2004.a} the
angularly dependent functions are expanded in a set of spherical harmonics, and
the assumption of translational invariance along $y$ allows us to write the
coefficients as functions of $\bm{s}$:
\begin{subequations}
\label{eqn:expansions}
\begin{align}
  \log\rho(\bm{r},\bm{u})&=\sum_{\ell,m}\lrho_{\ell m}(\bm{s})Y_{\ell m}(\bm{u}) \\
  \rho(\bm{r},\bm{u})&=\sum_{\ell,m}\rho_{\ell m}(\bm{s})Y_{\ell m}^*(\bm{u})             \\
  V_\text{ext}(\bm{r},\bm{u})&=\sum_{\ell,m}V_{\ell m}(\bm{s})Y_{\ell m}(\bm{u}) \;.
\end{align}
\end{subequations}
Note the complex conjugate in the density expansion. The
Mayer function is expanded as \cite{andrienko.d:2002.a}
\begin{equation}
  f(\bm{r}_{12},\bm{u}_1,\bm{u}_2)=\sum_{\ell_1,\ell_2,\ell}f_{\ell_1\ell_2\ell}(r_{12})
  \Phi_{\ell_1\ell_2\ell}(\hat{\bm{r}}_{12},\bm{u}_1,\bm{u}_2) \;,
\end{equation}
where $r_{12}=|\bm{r}_{12}|$, $\hat{\bm{r}}_{12}=\bm{r}_{12}/r_{12}$, and
$\Phi_{\ell_1\ell_2\ell}$ is a rotational invariant \cite{gray.cg:1984.a}.
Inserting these expressions into the grand potential and integrating over angles
and the $y$ direction gives
\begin{align}
  \frac{\beta\Omega[\rho(\bm{r},\bm{u})]}{L} &=
  \int\;\D \bm{s}\;\sum_{\ell,m}\rho_{\ell m}(\bm{s})\left(
    \lrho_{\ell m}(\bm{s})-\sqrt{4\pi}(1+\beta\mu)\delta_{l0}+\beta V_{\ell m}(\bm{s}) \right)
  \nonumber\\
  &+\int\;\D \bm{s}_1  \D \bm{s}_2 \sum_{\substack{\ell_1m_1\\ \ell_2m_2}}
  \mathcal{L}_{\ell_1m_1\ell_2m_2}(\bm{s}_{12})
  \rho_{\ell_1m_1}(\bm{s}_1)\rho_{\ell_2m_2}(\bm{s}_2) \;.
\label{eqn:grand}
\end{align}
Here $L$ is the box length in the $y$ direction (we assume periodicity). The
quantities $\mathcal{L}_{\ell_1m_1\ell_2m_2}(\bm{s}_{12})$ come from integrating
the Mayer function and are the spherical harmonic coefficients of the excluded
length (in the $y$ direction) of two molecules with a separation vector
$\bm{s}_{12}=\bm{s}_1-\bm{s}_2$ in the $xz$-plane, treated as a function of the
molecular orientations.  As the last term in eqn~\eqref{eqn:grand} is a
convolution, it is most conveniently evaluated in reciprocal space. If
$\rho_{lm}(\bm{k})$ is the two-dimensional Fourier transform of
$\rho_{lm}(\bm{s})$ then this term may be written
\begin{equation}
\sum_{\bm{k}}\sum_{\substack{\ell_1m_1\\ \ell_2m_2}}
  \mathcal{L}_{\ell_1m_1\ell_2m_2}(\bm{k})\rho_{\ell_1m_1}(\bm{k})
  \rho_{\ell_2m_2}(\bm{k})
\end{equation}
where $\mathcal{L}_{\ell_1m_1\ell_2m_2}(\bm{k})$ is the Fourier transform
of $\mathcal{L}_{\ell_1m_1\ell_2m_2}(\bm{s}_{12})$.

In order to find the equilibrium density the functions are tabulated on a
regular grid in the $xz$ plane; the grid spacing is $\delta x = \delta
z=0.2b$, the molecular length corresponding to 25 grid points. The grand
potential is then minimised with respect to the $\lrho_{lm}(\bm{s})$
coefficients at each grid point using the conjugate gradient method
\cite{press.wh:1986.a}. When required, the coefficients $\rho_{\ell
  m}(\bm{s})$ are calculated through eqns~\eqref{eqn:expansions}, with angular
integrations performed using Lebedev quadrature
\cite{lebedev.vi:1976.a,lebedev.vi:1977.a}. The systems in studied in this
work were all square boxes ($L_x=L_z$) with $L_x$ ranging from 40$b$ (200 grid
points) to 100$b$ (500 grid points). Table~\ref{tab:details} summarises the
parameters of the systems discussed in detail in the next section.
\begin{table}[hbt]
  \begin{tabular}{c c c c}
    \hline
    $R/b$  &  $L_x/b$  &  $n_x$  &  $\delta k_x/b^{-1}$   \\
    \hline
    5      &    40     &   200   &  $\pi/20$     \\
    7.5    &    60     &   300   &  $\pi/30$     \\
    10     &    60     &   300   &  $\pi/30$     \\
    15     &    80     &   400   &  $\pi/40$     \\
    20     &   100     &   500   &  $\pi/50$     \\
    \hline
  \end{tabular}
\caption{\label{tab:details} Details of systems studied. $R$ is the macroparticle radius, $L_x$ is
  the box length in the $x$ direction (with $L_z=L_x$), $n_x$  is the
  number of grid points in the $x$ direction and $\delta k_x$ is the
  grid spacing in $k$ space.}
\end{table}

Once the equilibrium density coefficients $\rho_{\ell m}(\bm{s})$ have been
determined, the number density $\rho(\bm{s})$ around the macroparticle may be
found from
\begin{equation}
  \rho(\bm{s})=\int \D\bm{u}\;\rho(\bm{s},\bm{u})=\sqrt{4\pi}\rho_{00}(\bm{s}).
\end{equation}
The orientational ordering is described by the order tensor
$Q_{\alpha\beta}(\bm{s})$ that is found from $\rho(\bm{s},\bm{u})$ by  
\begin{equation}
  Q_{\alpha\beta}(\bm{s})=
  \tfrac{3}{2}\int
  \D\bm{u}\;\rho(\bm{s},\bm{u})u_\alpha(\bm{s})u_\beta(\bm{s})- 
  \tfrac{1}{2}\delta_{\alpha\beta},\qquad\alpha,\beta=x,y,z \;.
\end{equation}
The spatially varying order parameter $S(\bm{s})$ is given by the
largest eigenvalue of $Q_{\alpha\beta}(\bm{s})$ and the director
$\bm{n}(\bm{s})$ by the eigenvector associated with  $S(\bm{s})$.

\section{Results}

\subsection{Structure in the nematic phase}

\begin{figure}[htp]
\includegraphics[width=0.8\textwidth,clip]{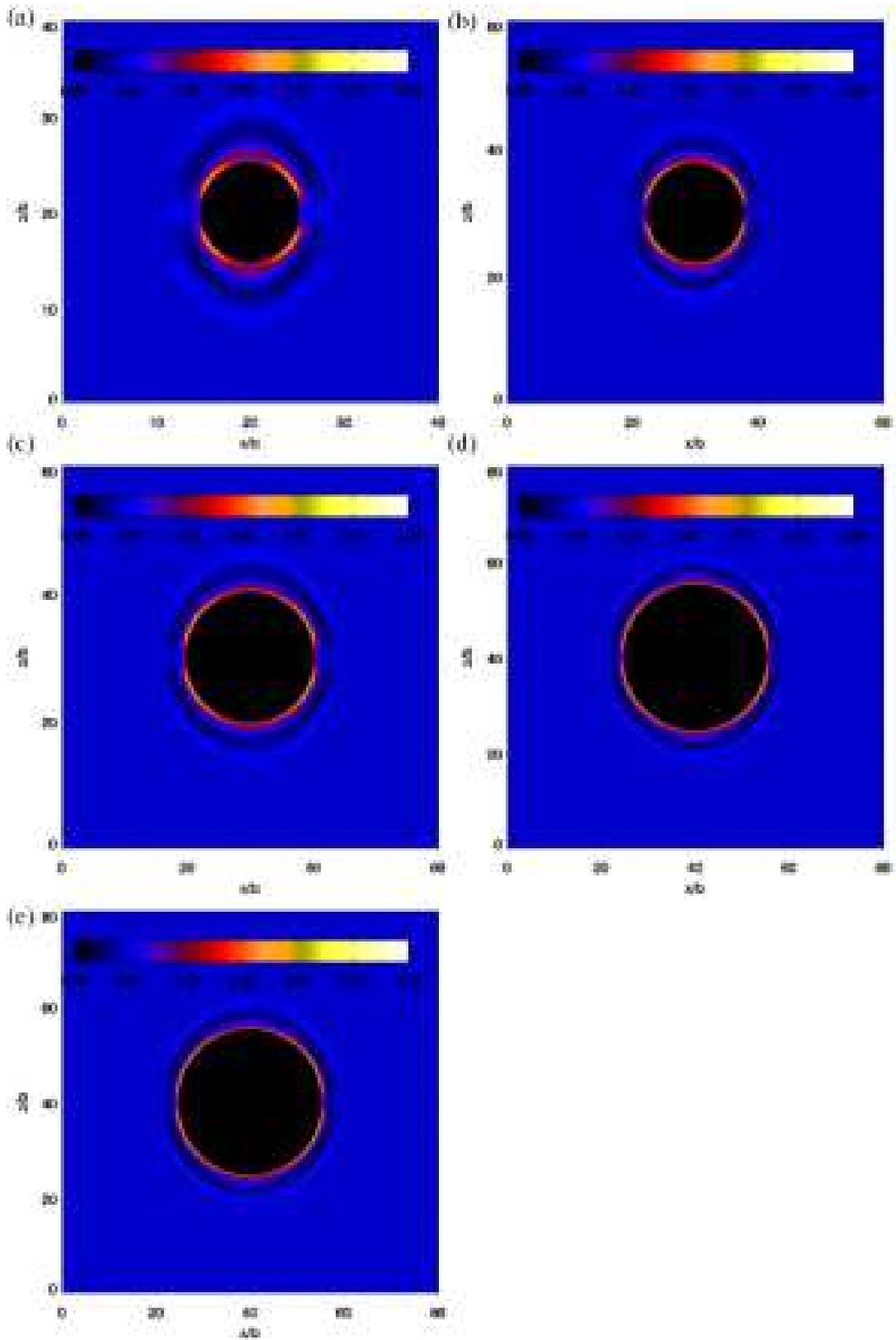}
\caption{\label{fig:den_map}%
  Density maps around a cylindrical macroparticle at $\mu=8.0$ (nematic phase).
  Cylinder radius $R/b$: (a) $5.0$, (b) $7.5$, (c) $10.0$, (d) $15.0$, and (e)
  $20.0$. Dark colours show areas of low $\rho(\bm{s})$, light
  colours high $\rho(\bm{s})$.}
\end{figure}

First we examine the fluid structure around the cylindrical particle in a
nematic fluid, at chemical potential $\mu=8.0$. The density distributions around
cylinders of radius $5\leq R/b \leq 20$ are shown in Fig.~\ref{fig:den_map}.
For all $R$ the density is largest at the surface but then decays away with
almost periodic variation, similar to that seen for a nematic-planar wall
interface.  Further from the cylinder this distortion in the density becomes
highly anisotropic.  Parallel to the director the modulations in the density are
stronger than in the perpendicular direction. This weakening perpendicular to
the director is due to the partial melting of the nematic in the defect regions
\cite{andrienko.d:2002.b} and is most noticeable for the smallest radius $R/b=5$
(Fig.~\ref{fig:den_map}a). In this case, there is almost no density modulation
perpendicular to the director causing the density map to show chevron-like
structures, with the chevron tips pointing along the director away from the axis
of the cylinder. For larger radii, the density variation perpendicular to the
director is stronger, but shorter-ranged, than in the direction parallel to the
director. In comparison to simulation of a system with $e=3$
\cite{andrienko.d:2002.b}, the density variation seems to be shorter ranged.
While this may be due to differences in the models, studies of the nematic-wall
interface using Onsager and related theories
\cite{cheung.dl:2004.a,chrzanowska.a:2001.a} also gave density profiles that
have generally weaker structure than comparable simulations.

\begin{figure}[htp]
\includegraphics[width=0.8\textwidth,clip]{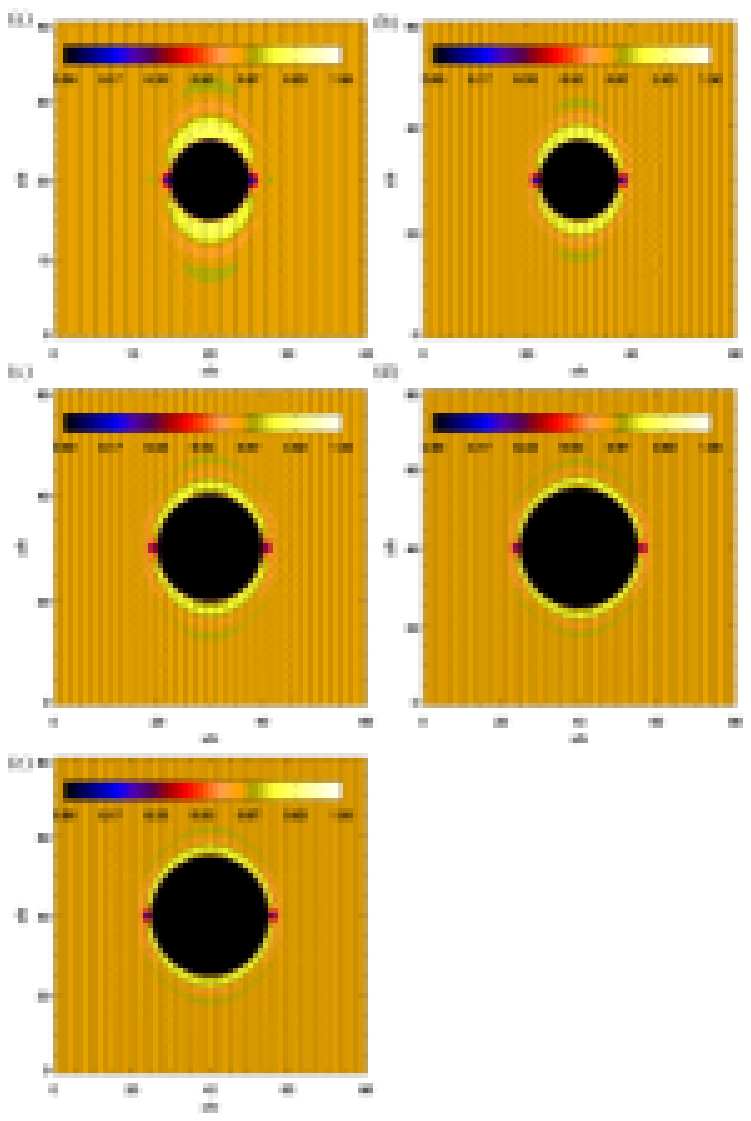}
\caption{\label{fig:order_map}%
  Order parameter maps around a cylindrical macroparticle at $\mu=8.0$ (nematic
  phase). Cylinder radius $R/b$: (a) $5.0$, (b) $7.5$, (c) $10.0$, (d) $15.0$,
  and (e) $20.0$. Dark colours show areas of low $S(\bm{s})$, light
  colours high $S(\bm{s})$. Solid lines show the orientation of the local
  director $\bm{n}(\bm{s})$.}
\end{figure}

Maps of the orientational order parameter around the cylindrical particles are
shown in Fig.~\ref{fig:order_map}. 
As for the density maps, the variation in the order parameter is highly
anisotropic. For $R/b=5$, along the director, there are lobes of high order
along the top and bottom of the cylinder. In the direction perpendicular to the
director there are two regions of drastically reduced order, corresponding to
defects in the liquid. As $R$ increases the lobes of increased order tend to
wrap around the cylinder, and the defects move away from the surface, in
agreement with simulation \cite{andrienko.d:2002.b} and phenomenological theory
\cite{burylov.sv:1994.a,fukuda.j:2001.a,tasinkevych.m:2002.a}. While the
positions of the defects are in qualitative agreement with previous results, in
the present case the defects are significantly smaller than those seen
previously, suggesting that for molecules of this elongation more sophisticated
methods such as weighted density approximations \cite{somoza.am:1989.a} or
fundamental measure theory \cite{cinacchi.g:2002.a} are necessary to examine the
structure within the topological defect. Detailed comparisons with simulation
will appear in a later publication.

Also shown in Fig.~\ref{fig:order_map} is the director orientation
$\bm{n}(\bm{s})$ around the macroparticle. At large distances from the
cylinder the $\bm{n}(\bm{s})$ lies along the $z$ axis. Close to the
particle the director becomes highly distorted. Along the $z$ axis and
at the defects $\bm{n}(\bm{s})$ is normal to the particle surface. At
other points on the particle surface, $\bm{n}(\bm{s})$ points away from
the surface normal appearing to graze the surface of the macroparticle. This behaviour is
different from that seen in simulation \cite{andrienko.d:2002.b} and
from elastic theory in the case of strong anchoring
\cite{burylov.sv:1994.a}, suggesting that the anchoring around the
macroparticle is weaker than in previous studies. Apart from in the
vicinity of the defects $\bm{n}(\bm{s})$ is a smoothly varying
function of $\bm{s}$.

\begin{figure}[htp]
\includegraphics[height=0.65\textwidth,clip]{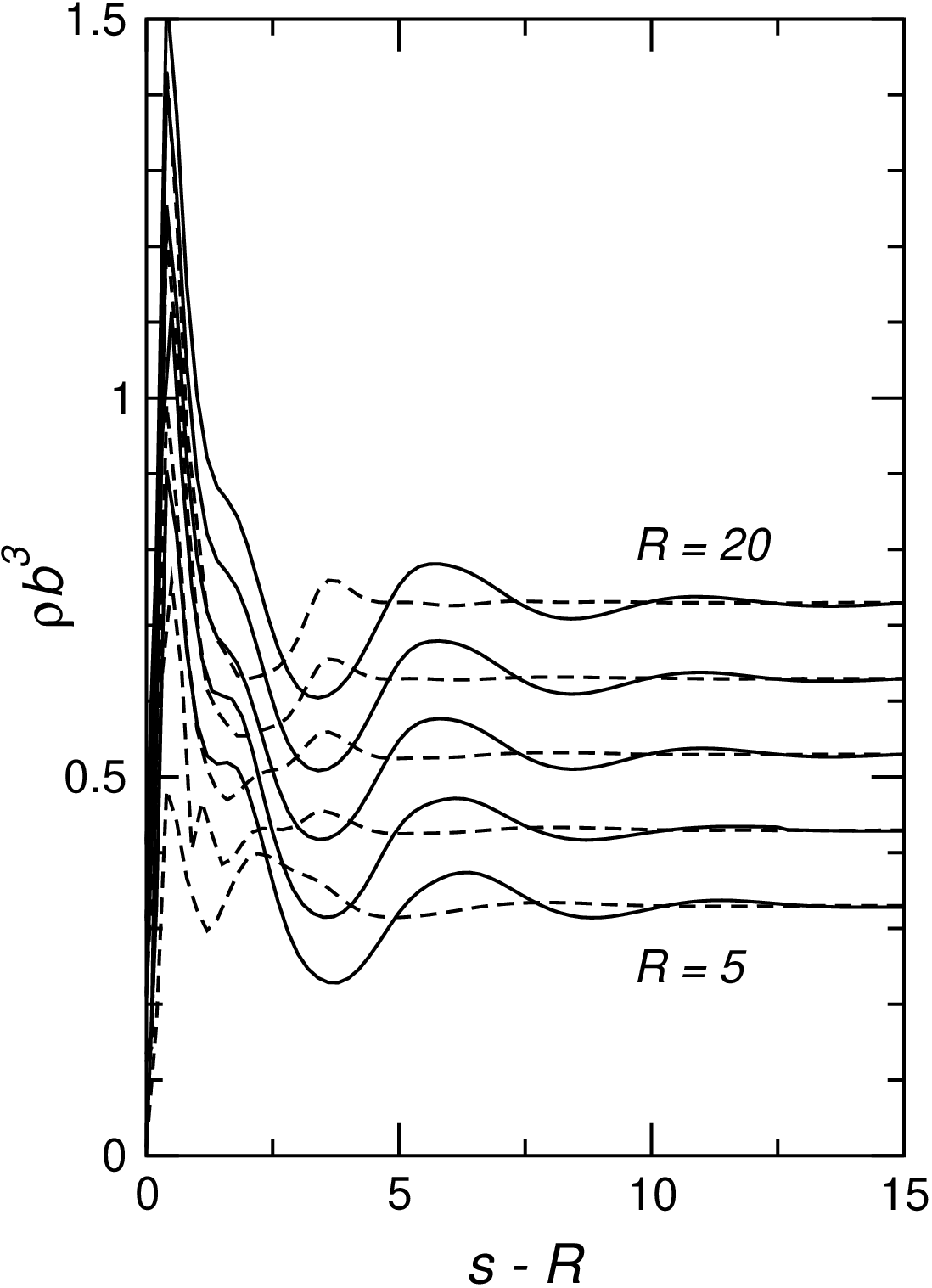}%
\hspace{\fill}
\includegraphics[height=0.65\textwidth,clip]{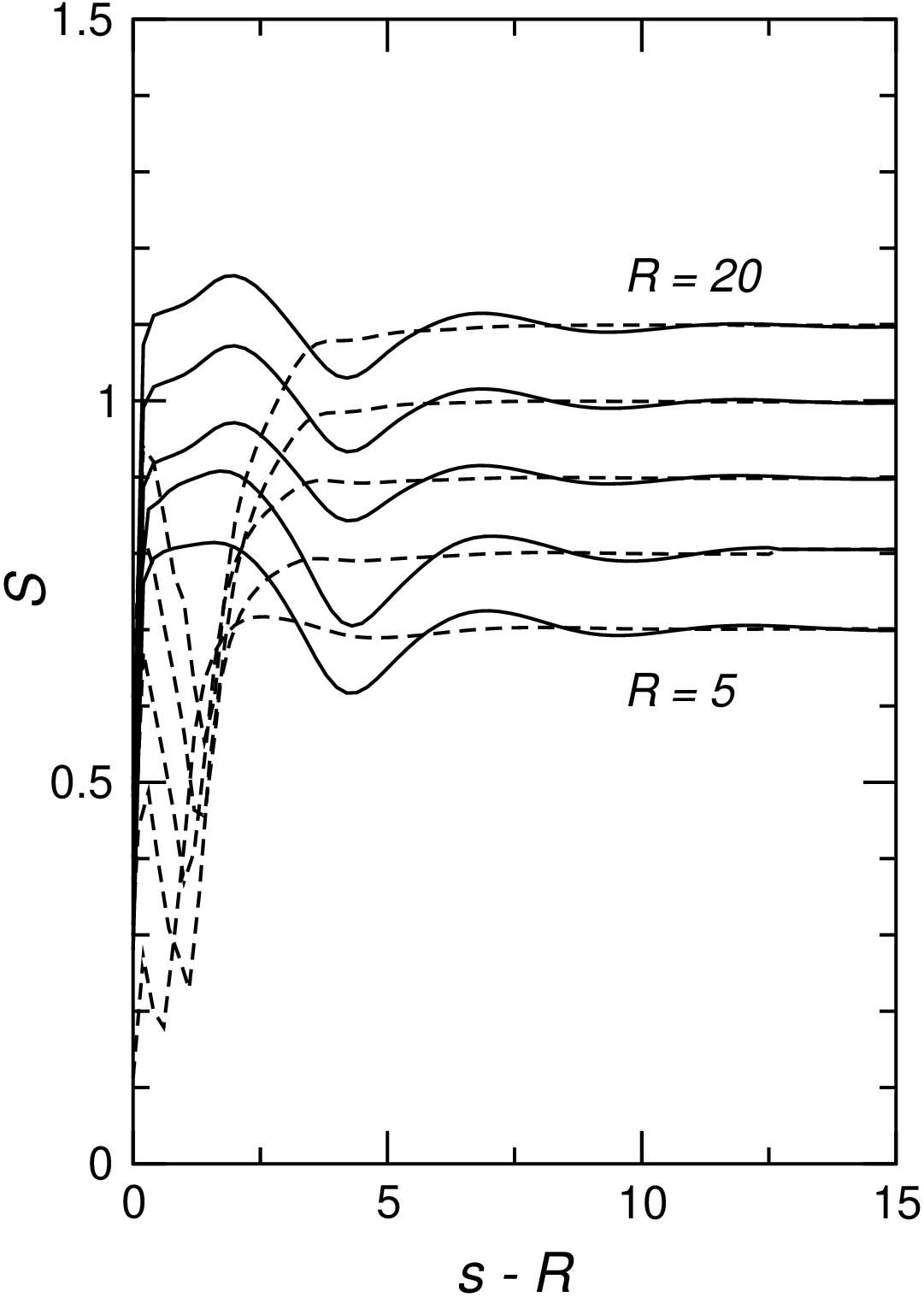}%
\caption{\label{fig:profiles}}
  (a) Density profiles, $\rho(s)b^3$, and (b) order parameter profiles $S(s)$,
  around a cylindrical macroparticle at $\mu=8.0$ (nematic phase). Full lines:
  in the direction parallel to the director. Dashed lines: in the direction
  perpendicular to the director. Results are shown for cylinder radius $R/b=5.0$
  and, displaced successively upwards by 0.1 units for clarity, $R/b=7.5, 10.0,
  15.0, 20.0$.
\end{figure}

In order to gain more insight, the density and order parameter profiles parallel
($\rho_\parallel(s)$ and $S_\parallel(s)$) and perpendicular ($\rho_\perp(s)$
and $S_\perp(s)$) to the director (through the defects) are shown in
Fig.~\ref{fig:profiles}.
The $\rho_\parallel(s)$ curves show similar structure as $R$ increases, with a
peak at the cylinder surface $s=R$, a second peak about $5b$ (one molecular
length) from the surface, and some decaying oscillations into the bulk. This is
similar to the behaviour seen for a nematic fluid near a wall with homeotropic
alignment \cite{allen.mp:1999.a,cheung.dl:2004.a}. The size of the peak at
contact increases with $R$ at small $R$. $\rho_\perp(s)$ also has a peak at
contact. For smaller cylinders $\rho_\perp(R)\ll \rho_\parallel(R)$. As $R$
increases the difference between the two heights decreases. Further from contact
$\rho_\perp(s)$ has a secondary peak. For the $R/b=5$ cylinder this peak is
approximately $2.5b$ from the surface with $\rho_\perp(s)$ tending towards the
bulk value further out.  For larger $R$, the secondary peak moves out. Beyond
this peak $\rho_\perp(s)$ behaves almost identically for all $R$.

\begin{figure}[htp]
\includegraphics[width=0.8\textwidth,clip]{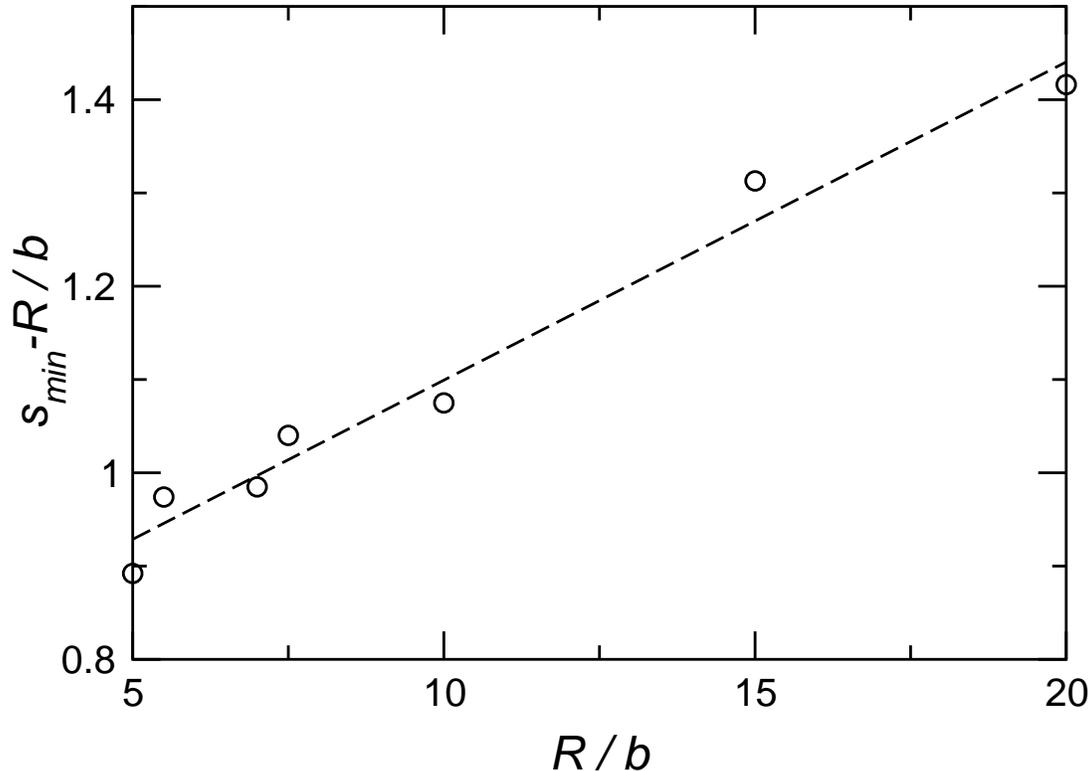}%
\caption{\label{fig:defect_pos}%
Defect position ($s_\text{min}-R$) near cylindrical macroparticle
at $\mu=8.0$ (nematic phase) as a function of radius $R$. Dashed line
shows the line of best fit.
}
\end{figure}

As with $\rho_\parallel(s)$, the $S_\parallel(s)$ profile is similar to the order
parameter profile for a nematic-homeotropic wall interface, with a maximum at
contact and decaying oscillations. $S_\perp(r)$ shows a minimum near contact.
This corresponds to the defect seen in the order parameter maps
(Fig.~\ref{fig:order_map}). In agreement with the order parameter maps, the
distance between the minimum and the surface increases with $R$. Shown
in Fig.~\ref{fig:defect_pos} is the position of the minimum $s_\text{min}$ as a
function of radius $R$.
For spherical macroparticles with both Saturn ring and
satellite defects, the distance between the surface and the defect varies
linearly with particle size \cite{andrienko.d:2001.a}. In the present case
$s_{min}$ increases with $R$.

\subsection{Structure in the isotropic phase and around the  nematic-isotropic transition}

\begin{figure}[htp]
\includegraphics[width=0.8\textwidth,clip]{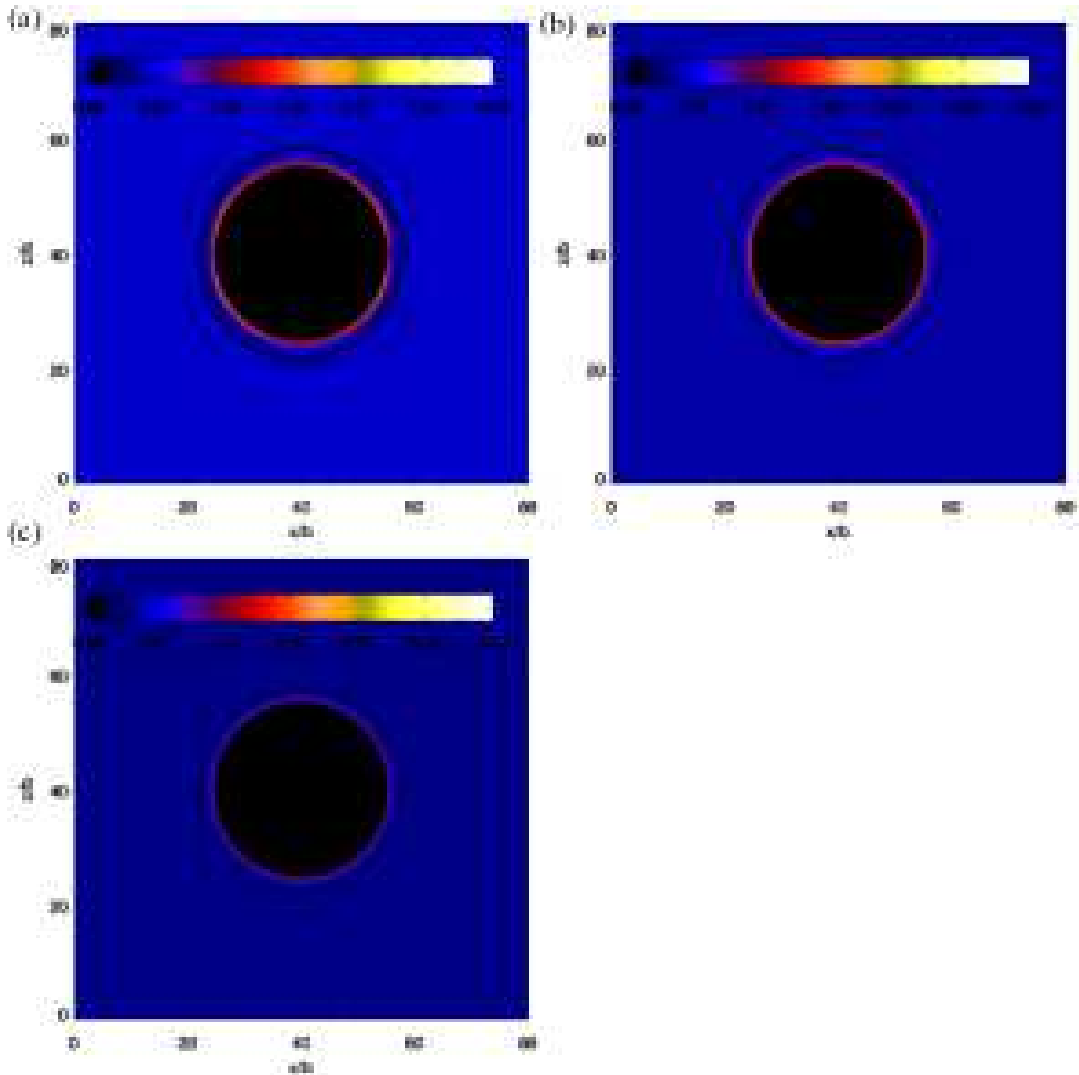}%
\caption{\label{fig:den_ni_map}%
  Density maps for fluid around cylindrical macroparticle of radius $R/b=15$
  with (a) $\mu=7.3675$ (nematic phase, at NI transition), (b) $\mu=7.3675$
  (isotropic phase, at NI transition), and (c) $\mu=5.0$ (isotropic
  phase). Dark colours show low $\rho(\bm{s})$, light colours high $\rho(\bm{s})$. }
\end{figure}

\begin{figure}[htp]
\includegraphics[width=0.8\textwidth,clip]{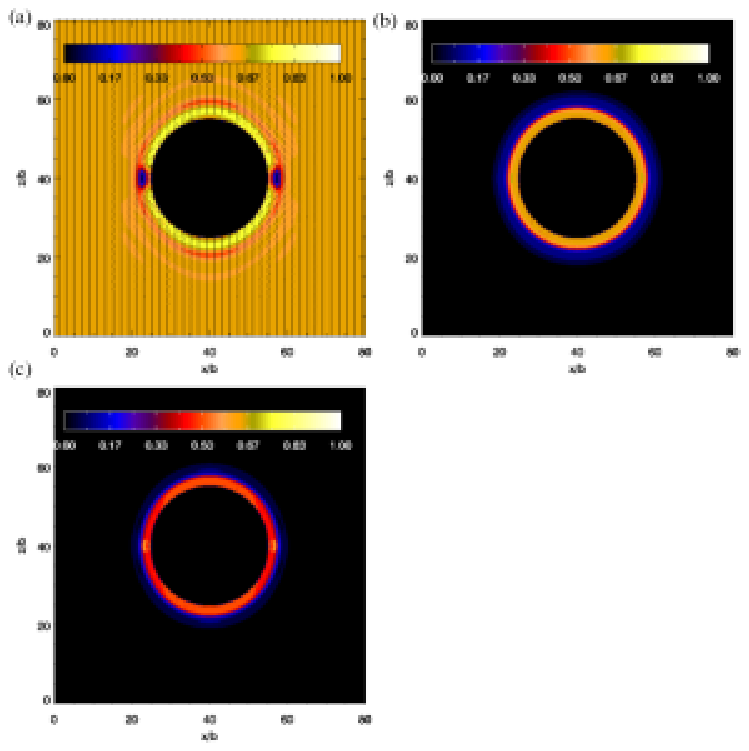}%
\caption{\label{fig:order_ni_map}%
  Order parameter maps for fluid around cylindrical macroparticle of radius
  $R/b=15$ with (a) $\mu=7.3675$ (nematic phase, at NI transition),
  (b) $\mu=7.3675$ (isotropic phase, at NI transition), and (c)
  $\mu=5.0$ (isotropic phase).  Dark colours show areas of low
  $S(\bm{s})$, light colours high $S(\bm{s})$.}
\end{figure}

Shown in Figs.~\ref{fig:den_map}d, \ref{fig:order_map}d, \ref{fig:den_ni_map}
and \ref{fig:order_ni_map} are the density and order parameter maps for a
cylindrical macroparticle with $R/b=15$ in both the nematic and isotropic
phases.
The bulk isotropic-nematic transition occurs at $\mu_{NI}=7.3675$. In
the nematic phase, at this value
of $\mu$, the density and order parameter maps are similar to those seen for $\mu=8$, though the
variation in the density and order parameter are less
pronounced. Also the defects either side of the cylinder with $\mu=7.3675$ are both larger and significantly further
from the surface ($s_\text{min}-R\approx2.5b$) than for the higher chemical
potential. In the isotropic phase, both the density and order parameter
variation become almost completely symmetrical, reflecting the loss of a
preferred direction. In this phase, far from the cylinder, the order parameter
$S\rightarrow0$, while both $\rho$ and $S$ have maxima at the surface. This
suggests that the surface is wet by the nematic and this gives rise to short
range interactions between particles dispersed in a liquid crystal host even in
the isotropic phase \cite{borstnik.a:1999.a,kocevar.k:2001.a}. However, a
detailed investigation of surface phase behaviour and interparticle interactions
in this model await further study.

\section{Conclusions}

In this paper the structure of a liquid of hard ellipsoids of elongation $e=5$,
around a cylindrical macroparticle, in both the nematic and isotropic phases,
was studied using density functional theory within the Onsager approximation.
The resulting density and order parameter maps were consistent with previous theoretical and simulation work. On going from the
nematic to isotropic phase, the structure of the surrounding fluid becomes
rotationally invariant about the cylinder, with what appears to be a nematic
wetting layer at the particle surface.

The present study is preliminary: investigation of the sensitivity of the
results to the resolution of the real-space and reciprocal-space grids must
still be carried out.  Nonetheless, the results are very promising. Despite its
simplicity, this method provides results that are generally in agreement with
simulation and phenomenological theory. The only major deficiency is that the
defects are smaller than would be expected from simulations of a similar system
with $e=3$. Using more sophisticated density functionals or ellipsoids with
longer elongations (where Onsager theory is more accurate) would hopefully give
a better description of the defect. Detailed comparisons with simulation will be
described elsewhere.

\section*{Acknowledgments}

This research was supported by EPSRC grant GR/S77240.
The calculations were performed on the computing facilities of the
Centre for Scientific Computing, University of Warwick.
The authors of Ref.~\cite{harnau.l:to_appear} are thanked for providing a preprint.

\end{document}